\documentclass[doublecol,linenumbers]{epl2}

\usepackage{graphics}% Include figure files
\usepackage{amsmath,amssymb,mathtools}
\usepackage[english]{babel}
\usepackage{epstopdf}
\usepackage[caption=false]{subfig}
\usepackage{color}
\DeclareMathOperator\erf{erf}

\title{Conditional entropy  and Landauer principle}
\shorttitle{Conditional entropy  and Landauer principle}

\author{D. Chiuchi\'u\inst{1}\thanks{E-mail: \email{davide.chiuchiu@nipslab.org}} \and M. C. Diamantini\inst{1,2} \and L. Gammaitoni\inst{1}}
\shortauthor{D. Chiuchi\'u\etal}

\institute{                    
  \inst{1} NiPS Laboratory, Dipartimento di Fisica, Universit\'a degli studi di Perugia - Via Alessandro Pascoli, I-06123 Perugia, Italy\\
  \inst{2} INFN, sezione Perugia - I-06100 Perugia, Italy
}

\abstract{Landauer principle describes the minimum heat produced by an information-processing device. Recently a new term has been included in the minimum heat production: it's called  conditional entropy and takes into account the microstates content of a given logic state. Usually this term is assumed zero in bistable symmetric systems thanks to the strong hypothesis used to derive Landauer principle. In this paper we show that conditional entropy can be nonzero even for bistable symmetric systems and that it can be expressed as the sum of three different terms related to the probabilistic features of the device. The contribution of the three terms to conditional entropy (and thus to minimum heat production) is then discussed for the case of bit-reset.}

\begin{document}

\maketitle

\section{I. Introduction}

Landauer principle relates information theory and thermodynamics\cite{landauer}: it express a quantitative relation for the minimum heat produced by a computing device during information processing. Its simplest formulation states that at least $K_bT \ln 2$ of heat has to be produced to reset one bit in binary symmetric systems. Recently, a more sophisticated formulation has been proposed \cite{sagawa,maroney} that takes into account the role of conditional entropy \cite{cover}  to relate Shannon and Gibbs Entropy. In these works the role of conditional entropy has been emphasized in the presence of asymmetric bistable systems.
In this paper we show that asymmetry is not a necessary condition and that conditional entropy can significantly affect the minimum heat production even for symmetric devices. To see this point, we consider a computing device as generic physical system with an Avogadro number $N$ of degrees of freedom (DOF). Those are classified in two categories: one DOF is chosen to characterize the relevant system dynamic and the computation process, while the remaining $N -1$ DOFs behave as a thermal bath at constant temperature $T$. The dynamic-relevant DOF is labelled $x$ and, by construction, it can assume a large number of values, possibly continuum, each corresponding to one different microstate of the system. The set of possible $x$ values, $\Omega$, is the ensemble of all the microstates of the device. Because of the thermal bath, $x$ fluctuates and we can define $P(x)$, the probability density function (PDF) of $x$. 
The thermodynamic Gibbs entropy \cite{jaynes,seifert} of the system is then defined as: 
\begin{equation}\label{eq:gibbsentropy}
S_{G}=-K_b\int_{\Omega}P(x)\ln{P(x)}\;\upd x
\end{equation}
where $K_b$ is Boltzmann constant. 
\\To encode $n$ different logic states in the device, $\Omega$ is divided in $n$ non-overlapping subsets $\Omega_0\ldots \Omega_{n-1}$, each one containing the microstates consistent with the given logic state\footnote{There is no unique way to make such division and the particular choice for $\Omega_0\ldots\Omega_{n-1}$ is application dependent.}. The probability of assuming the $i$-th logic state is
\begin{equation}\label{eq:probs}
P_i=\int_{\Omega_i}P(x)\;\upd x
\end{equation}
and the information content of the device is described by Shannon information  entropy \cite{shannon,cover}
\begin{equation}\label{eq:shannonentropy}
S_S=-K_b\sum_{i=0}^{n-1} P_i\ln{P_i}.
\end{equation}

Gibbs and Shannon entropy are quite similar, however eq.(\ref{eq:probs}) implies that $S_S$ is a coarse grained version of $S_G$ where the number of internal microstates in each logic state is ignored. 
The relation between Gibbs and Shannon entropies is provided by \cite{maroney,sagawa}
\begin{equation}\label{eq:uguaglianzacorretta}
S_{G}=S_{S}+S_{cond}
\end{equation} 
where $S_{cond}$ is the contribution of the different microstates inside each $\Omega_i$.

Computation can be seen as an information manipulation performed through a physical transformation of the system. This physical transformation obeys Clausius theorem, $Q\geq-T\Delta S_{G}$, where $Q$ is the amount of heat exchanged with the reservoir\footnote{By convention, $Q$ is positive for heat given to the reservoir}. Using eq.(\ref{eq:uguaglianzacorretta}) we obtain
\begin{equation}\label{eq:Clausiuscorretto}
Q\geq -T\Delta S_{S}-T\Delta S_{cond}.
\end{equation} 
Eq.(\ref{eq:Clausiuscorretto}) is the generalized Landauer principle \cite{sagawa} where we can recognize a minimum heat production due to information processing ($-T\Delta S_S$) that can be increased or compensated by the entropy change inside each logic state ($-T\Delta S_{cond}$). 

Up to now, $\Delta S_{cond}=0$ was explicitly or implicitly assumed for any logic operation performed on bistable symmetric systems \cite{landauer,lambson,jarz, shizume, piechocinska,dillenlutz}, while $\Delta S_{cond}\neq0$ was considered possible only for asymmetric systems \cite{sagawa}. In this paper we show that $\Delta S_{cond}\neq 0$ is also valid  for binary symmetric systems and that it contributes significantly to minimum heat production. To prove this fact, we consider a bistable symmetric system and a continuous two-peaked PDF that generalizes the one used in \cite{sagawa,shizume, piechocinska,dillenlutz,ciliberto}. With this PDF we show that $\Delta S_{cond}$ can be expressed as the sum of three different terms connected to the PDF structure. The contribution of the three terms to conditional entropy (and thus to minimum heat production) is then discussed for the case of bit-reset.

\section{II. Conditional entropy as the sum of three contributions}

We assume that our microstates space $\Omega=]-\infty, \infty[$ is split in two subsets $\Omega_0=]-\infty,0[$ and $\Omega_1=]0,\infty[$. The physical system is said to encode one bit of information in the $0$ or $1$ logic state if $x\in\Omega_0$ or $x\in\Omega_1$, respectively. Probabilities of being in the $0$ or $1$ logic state and Shannon entropy are given by eq.(\ref{eq:probs}) and eq.(\ref{eq:shannonentropy}) with $n=2$ and $i=0,1$. 

To practically confine $x$ values within a given working range we introduce a static bistable and symmetric potential $U(x)$ with two minima in $x_a$ and $x_b$ separated by an energy barrier $\Delta U$ in $x=0$. Since the boundary between $\Omega_0$ and $\Omega_1$  lies at the top of the energy barrier,   $x_a$ and $x_b$ are contained in different logic subsets and the energy barrier guarantees state stability for a time shorter than the residence time \cite{gammaitonimarchesoni}. 

If the system is at thermal equilibrium with the reservoir, the associated PDF is given by the canonical distribution \cite{vanvliet}:
\begin{equation}\label{eq:thermaleq}
P(x)=\mathcal{Z}^{-1}e^{-\frac{U(x)}{K_bT}}
\end{equation}
where $\mathcal{Z}^{-1}$ is the partition function. Eq.(\ref{eq:thermaleq}) describes a system with equal probability $1/2$ of being either in the logical state 0 or in the logical states 1.
The corresponding non-equilibrium distribution is usually \cite{piechocinska, sagawa,dillenlutz} represented as:
\begin{equation}\label{eq:thermaleqproto}
P(x)=P_0\ 2\mathcal{Z}^{-1} e^{-\frac{U(x)}{K_bT}}\Theta(x)+P_1\ 2\mathcal{Z}^{-1} e^{-\frac{U(x)}{K_bT}}\Theta(-x)
\end{equation}
where $\Theta(x)$ is Heaviside step function, with $P_0$ and $P_1$  probabilities of being in the 0 and 1 state, arbitrarily adjusted. Results concerning Landauer limit and conditional entropy are then derived from eq.(\ref{eq:thermaleqproto}).

In this paper we assume a different PDF, that includes eq.(\ref{eq:thermaleqproto}) as a special case, but avoids its discontinuity when $P_0$ and $P_1$ differ from $1/2$. This is not just a mathematical nuisance: for any bistable system that left to reach a local equilibrium condition near its minima, the Fokker-Planck equation \cite{gardiner, risken} provides continuous $P(x)$ for continuous $U(x)$. Most importantly, our approach, at difference with previous results \cite{sagawa}, shows that $\Delta S_{cond}$ can be different from 0 even for symmetric bistable potentials if the height of the energy barrier becomes comparable with the thermal noise ($\Delta U\approx K_bT$).

To identify a more appropriate form of the PDF, we note that any bistable system with a continuous potential and a moderate noise intensity, will generally have $P(x)$ peaked around $x_a$ and $x_b$ and suppressed near the barrier. Based on these considerations we introduce two functions, $\eta_a(x)$ and $\eta_b(x)$, single-peaked with maximum approximately at $x_a$ and $x_b$ respectively. Their supports are labeled $\Omega_a$ and $\Omega_b$ and those may not coincide with $\Omega_0$ and $\Omega_1$. This additional freedom in $\Omega_a$ and $\Omega_b$ gives the advantage to choose $\eta_a$ and $\eta_b$ without discontinuities of the first kind as in eq.(\ref{eq:thermaleqproto}). To add to the generality, we allow  $\Omega_a$ and $\Omega_b$ to superimpose over a subset called $\Omega_{ov}$. We finally assume normalized functions $\int_{\Omega_{a,b}}\eta_{a,b}(x)\upd x=1$. In this way we can express the nonequilibrium PDF as: 
\begin{equation}\label{eq:formalPDF}
P(x)=P_a\eta_a(x)+P_b\eta_b(x).
\end{equation}
with
\begin{equation}\label{eq:abboundary}
P_a+P_b=1.
\end{equation}
where $P_a$ ($P_b$) describe the average probability that one microstate belongs to $\eta_a$ ($\eta_b$).
Application of eq.(\ref{eq:probs}) to eq.(\ref{eq:formalPDF}) gives 
\begin{equation}\label{eq:formalProbabilities}
\begin{aligned}
P_0&=P_a\smashoperator{\int_{\Omega_{a}\cap\Omega_0}}\eta_a(x) \upd x+P_b\smashoperator{\int_{\Omega_{b}\cap\Omega_0}}\eta_b(x) \upd x\\
P_1&=P_a\smashoperator{\int_{\Omega_{a}\cap\Omega_1}}\eta_a(x) \upd x+P_b\smashoperator{\int_{\Omega_{b}\cap\Omega_1}}\eta_b(x) \upd x
\end{aligned}
\end{equation}
showing a linear relationship between $P_0$, $P_1$, $P_a$ and $P_b$. We stress that eq.(\ref{eq:formalPDF}), (\ref{eq:abboundary}) and (\ref{eq:formalProbabilities}) are more general than eq.(\ref{eq:thermaleqproto}) and are valid even if $U(x)$ is simply bistable but not symmetric.

We now calculate Gibbs and conditional entropy using the PDF given in eq.(\ref{eq:formalPDF}). Gibbs entropy is given by:
\begin{equation}
\begin{aligned}
\frac{S_G}{K_b}=&-\int_{\Omega}(P_a\eta_a+P_b\eta_b)\ln{(P_a\eta_a+P_b\eta_b)}\upd x=\\
=&-\smashoperator{\int_{\Omega_a\setminus \Omega_{ov}}}P_a\eta_a\ln{(P_a\eta_a)}\upd x -\smashoperator{\int_{\Omega_b\setminus \Omega_{ov}}}P_b\eta_b\ln{(P_b\eta_b)}\upd x+\\
&-\int_{\Omega_{ov}}P_a\eta_a\ln{\bigg(P_a\eta_a\Big(1+\frac{P_b}{P_a}\frac{\eta_b}{\eta_a}\Big)\bigg)}\upd x+\\
&-\int_{\Omega_{ov}}P_b\eta_b\ln{\bigg(P_b\eta_b\Big(1+\frac{P_a}{P_b}\frac{\eta_a}{\eta_b}\Big)\bigg)}\upd x=\\
=&-\int_{\Omega_a}P_a\eta_a\ln{(P_a\eta_a)}\upd x -\int_{\Omega_b}P_b\eta_b\ln{(P_b\eta_b)}\upd x+\\&+\frac{P_a}{K_b}I\big(\eta_a,\eta_b,\tfrac{P_b}{P_a} \big)+\frac{P_b}{K_b}I\big(\eta_b,\eta_a,\tfrac{P_a}{P_b} \big)
\end{aligned}
\end{equation}
 with
\begin{equation}\label{eq:formalOverlap}
I(\eta_a,\eta_b,q)=-K_b\int_{\Omega_{ov}}\eta_a\ln{\bigg(1+q \frac{\eta_b}{\eta_a} \bigg)}\upd x
\end{equation}
After some additional calculations and the definition of Gibbs single-peak entropy $S_a=-K_b\int_{\Omega_a}\eta_a\ln\eta_a\upd x$, $S_G$ can be expressed as the sum of three terms: 
\begin{equation}\label{eq:formalGibbs}
S_G = S_{cg}+S_{pe}+S_{ov}
\end{equation}
with
\begin{subequations}\label{eq:formalGibbs3}
\begin{align}
S_{cg}=&-K_b(P_a\ln{P_a}+P_b\ln{P_b})\\
S_{pe}=&P_aS_a+P_bS_b\\
S_{ov}=&P_aI\big(\eta_a,\eta_b,\tfrac{P_b}{P_a}\big)+P_bI\big(\eta_b,\eta_a,\tfrac{P_a}{P_b}\big).
\end{align}
\end{subequations}

The first is $S_{cg}$, a coarse grained entropy. It is the Shannon entropy built with probabilities $P_a$, $P_b$ that a microstate belongs to $\eta_a$ and $\eta_b$. $S_{pe}$ describes entropy contributions arising from the shapes of the two peaks when those are considered non-overlapped. Corrections due to overlap are given in $S_{ov}$ by the $I$ integrals.  

Conditional entropy $S_{cond}$ is then given as three contributions as
\begin{equation}\label{eq:formalConditional}
S_{cond}=S_G-S_S=S_{ex}+S_{pe}+S_{ov}
\end{equation}
with
\begin{equation}
S_{ex}=S_{cg}-S_{S}.
\end{equation}
Here $S_{pe}$ and $S_{ov}$ are the same as in eq.(\ref{eq:formalGibbs}). The remaining term, $S_{ex}$, is the difference between the entropy eq.(\ref{eq:formalGibbs3}a) (describing microstate coarse graining in two peaks) and Shannon entropy (describing microstate coarse graining in two logic subsets). It thus represents the absolute entropic measure of the error we commit if we  exchange probability set $(P_a, P_b)$ with $(P_0,P_1)$. Conditional entropy variation is given by:
\begin{equation}\label{eq:deltaScondFormal}
\Delta S_{cond}=\Delta S_{ex}+\Delta S_{pe}+\Delta S_{ov}.
\end{equation}
Here $\Delta S_{ex}$  takes into account the possible change of the value of $P_a$ and $P_b$,   and the fact that  peaks can move from one logic subset to the other during the transformation (this reflects the change in the overlap between $\Omega_a$ and $\Omega_b$ with $\Omega_0$ and $\Omega_1$). $\Delta S_{pe}$ arises from the changes in shape of the peaks, while  $\Delta S_{ov}$ from the change in the way the two peaks overlap.
\\In the next sections we consider two examples to better illustrate eq.\ref{eq:deltaScondFormal}.

\section{III. First example}
As first example, we recall that eq.(\ref{eq:thermaleqproto}) is eq.(\ref{eq:formalPDF}) with 
\begin{subequations}\label{eq:caso1}
\begin{gather}
\eta_a=2\mathcal{Z}^{-1} \exp(-U(x)/(K_bT))\Theta(x)\\
\eta_b=2\mathcal{Z}^{-1} \exp(-U(x)/(K_bT))\Theta(-x)\\
\Omega_a=\Omega_0,\quad \Omega_b=\Omega_1
\end{gather}
\end{subequations}
Conditional entropy is easily computed. Eq.(\ref{eq:caso1}c) and eq.(\ref{eq:formalProbabilities}) gives that $P_0=P_a$ and $P_1=P_b$ so $S_{cg}=0$. Also $S_{ov}=0$, because $\Omega_{ov}=\Omega_0\cap\Omega_1=\varnothing$. As a consequence
\begin{equation}
S_{cond}=S_{pe}=-K_b\ln{2}+K_b\Big(\frac{\langle U\rangle}{K_bT} +\ln{\mathcal{Z}^{-1}}\Big)
\end{equation}
where $\langle U\rangle$ is the average value of $U(x)$ at thermal equilibrium. 

Now, when eq.(\ref{eq:thermaleqproto}) is used as nonequilibrium PDF \cite{piechocinska, shizume, dillenlutz}, logic operations are isothermal transformations that change $(P_0,P_1)$ from some initial values $(P_0^i,P_0^i)$ into some others $(P_0^f,P_1^f)$. As a consequence $\Delta S_{cond}=0$ for any logic operation. However, this is not a general property of symmetric bistable systems \cite{sagawa} but a consequence of the assumption of eq.(\ref{eq:thermaleqproto}) as nonequilibrium PDF. We illustrate this point with the second example. 

\section{IV. Gaussian example}

Let us consider the system with symmetric bistable potential $U(x)$  at thermal equilibrium with a reservoir at a temperature $T$.
Due to the symmetry $\eta_b(x)=\eta_a(-x)$ and
\begin{equation}\label{eq:initialProbabilitiesab}
P_a^i=P_b^i=\frac{1}{2}.
\end{equation}
From eq.(\ref{eq:formalProbabilities}) we have
\begin{equation}\label{eq:initialProbabilities}
P_0^i=P_1^i=\frac{1}{2}
\end{equation}
i.e. the initial state of the bit is undefined.

A simple analytical expression for $\eta_a(x)$ can be obtained through harmonic approximation of $U(x)$ near $x_a$ minimum in eq.(\ref{eq:thermaleq}):
\begin{equation}\label{eq:gausspeak}
\eta_a(x)\approx\mathcal{N}e^{-\frac{(x+h\sigma)^2}{2\sigma^2}}
\end{equation}
with $\sigma$ fitted from local harmonic approximation 
\begin{equation}\label{eq:sigmadef}
\sigma=\sqrt{\frac{K_bT}{ U^{\prime\prime}(x_a) }},
\end{equation}
$U^{\prime\prime}$ the second derivative respect to $x$, $\mathcal{N}=(\sqrt{2\pi}\sigma)^{-1}$ and
\begin{equation}\label{eq:tdef}
h=\frac{|x_a|}{\sigma}=\sqrt{2}\sqrt{\frac{ \frac{1}{2} U^{\prime\prime}(x_a)  x_a^2 }{K_bT}}.
\end{equation}
The numerator of eq.(\ref{eq:tdef}) is the harmonically approximated potential energy of a point distant $|x_{a}|$ from $x_a$. If $U(x)$ is smooth, this quantity has approximately the same amplitude of $\Delta U$, implying $h\approx [2\Delta U/(K_bT)]^{\frac{1}{2
}}$ and
\begin{equation}\label{eq:tedeltaU}
h\gg 1 \equiv \frac{\Delta U}{K_b T}\gg1
\end{equation}

A finite-time protocol is then implemented to reset the logic state to zero, i.e. the average value of $x$ becomes negative. The protocol may be the same given in \cite{ciliberto}, but any physical transformation that recovers $U(x)$ at the end and that makes the average value of $x$ negative is indeed a valid choice. If both $U(x)$ and the protocol are specified, then the final PDF is solution of the associated Fokker-Planck equation \cite{gardiner,risken}. The main limitation of this approach is that analytical solutions are difficult to obtain. For this reason we assume that at the end of the protocol the system is allowed to reach a local equilibrium condition around the minima of $U(x)$. The final PDF is then correctly approximated by eq.(\ref{eq:formalPDF}) with $\eta_b(x)=\eta_a(-x)$ and $\eta_a$, $\sigma$ and $h$ given by eq.(\ref{eq:gausspeak}), (\ref{eq:sigmadef}) and (\ref{eq:tdef}). The sole difference with the initial distribution is that $P_a^f$ can take any value in $[0.5,1]$ so to have negative $x$ average value. In this way we can also study reset protocols where there is a non-negligible probability $P^f_1$ to end in the wrong logic state. Putting eq.(\ref{eq:gausspeak}), $\eta_b(x)=\eta_a(-x)$ and $\Omega_a=\Omega_b=\Omega=]-\infty,\infty[$ in eq.(\ref{eq:formalProbabilities}), after some algebraic manipulations 
we obtain 
\begin{equation}\label{eq:finalProbabilities}
\begin{aligned}
P_0^f&= \frac{1+(2P_a^f-1)\erf(\frac{h}{\sqrt{\pi}})}{2}\\
P_1^f&= \frac{1+(2P_b^f-1)\erf(\frac{h}{\sqrt{\pi}})}{2}.
\end{aligned}
\end{equation}
where $\erf(y)=2 \pi^{-\frac{1}{2}} \int_0^y e^{-z^2} dz $ is the error function.

\begin{figure}
\includegraphics[width=\linewidth]{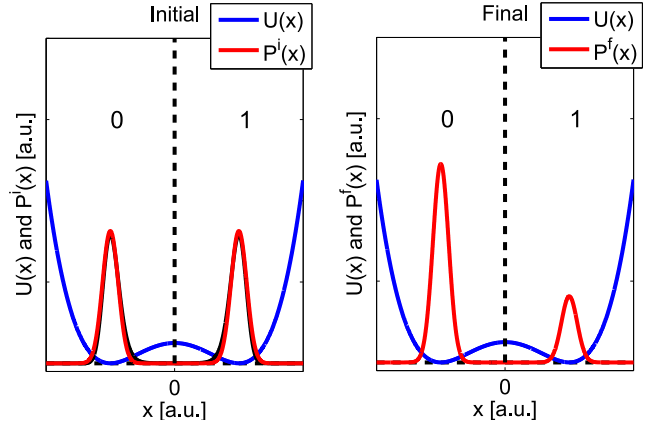}
\caption{Schematic representation of a sample reset protocol. Left and right plots describe the initial and final state of the bit. Potenial $U(x)$ (blue) is the same at the begin and the end of the protocol. Initial and final distributions are shown in red. Final distribution has a negative average value of $x$, so the bit is on the average resetted to zero.}\label{fig:figure2}
\end{figure}

Summarizing, we have that our bit-encoding system is initially at thermal equilibrium: its initial PDF (Fig. \ref{fig:figure2}, left) is $P(x) = \eta_a(x)/2+\eta_a(-x)/2$ with $\eta_a(x)$ given by eq.(\ref{eq:gausspeak}). Logic states probabilities are given by eq.(\ref{eq:initialProbabilities}). We then perform a finite-time reset to zero protocol which brings the system in the final PDF (Fig. \ref{fig:figure2}, right) described by eq.(\ref{eq:formalPDF}) with $\eta_b(x)=\eta_a(-x)$, $\eta_a(x)$ given by eq.(\ref{eq:gausspeak}) and $P_a^f\in[0.5,1]$. Logic state state probabilities are given by eq.(\ref{eq:finalProbabilities}). The hypothesis here used are common in experiments on bit reset \cite{ciliberto,ciliberto2,lambson} with the sole new addition that initial and final probability density functions have the form of eq.(\ref{eq:formalPDF}). This choice is justified for reset protocols which brings the system at least in local equilibrium near potential minima. The possibility of reset errors is also included in this description.

Next step is the computation of the entropy variations from their definitions with the Gaussian peaks. Gibbs entropy, $\Delta S_{G}$ is given by:
\begin{equation}\label{eq:deltaSGibbs}
\begin{aligned}
\frac{\Delta S_G}{K_b}=&-P_a^f\ln{P_a^f}-P_b^f\ln{P_b^f} -\ln{2}\\
&-\frac{P_a^f}{\sqrt{2\pi}}\int_{-\infty}^\infty e^{-\frac{z^2}{2}}\ln\bigg(1+\frac{P_b^f}{P_a^f} e^{2zh-2h^2}\bigg)dz+\\
&-\frac{P_b^f}{\sqrt{2\pi}}\int_{-\infty}^\infty e^{-\frac{z^2}{2}}\ln\bigg(1+\frac{P_a^f}{P_b^f} e^{2zh-2h^2}\bigg)dz+\\
&+\frac{1}{\sqrt{2\pi}}\int_{-\infty}^\infty e^{-\frac{z^2}{2}}\ln\bigg(1+e^{2zh-2h^2}\bigg)dz
\end{aligned}
\end{equation}
\normalsize 
with $z=(x\pm h\sigma)/\sigma$. 
The first line of the RHS represents $\Delta S_{cg}$, while the last three lines are $\Delta S_{ov}$. 
$\Delta S_{pe}=0$ here because eq.(\ref{eq:abboundary}) is satisfied and $\eta_a(x)$ is the same at the beginning and the end of the protocol.

Shannon entropy variation is promptly obtained through application of eq.(\ref{eq:initialProbabilities}) and (\ref{eq:finalProbabilities}) to eq.(\ref{eq:shannonentropy})
\small
\begin{equation}\label{eq:deltaSShannon}
\begin{aligned}
\frac{\Delta S_S}{K_b}=&-\frac{1+(2P_a^f-1)\erf(\frac{h}{\sqrt{\pi}})}{2}\ln\frac{1+(2P_a^f-1)\erf(\frac{h}{\sqrt{\pi}})}{2}+\\ &-\frac{1+(2P_b^f-1)\erf(\frac{h}{\sqrt{\pi}})}{2}\ln\frac{1+(2P_b^f-1)\erf(\frac{h}{\sqrt{\pi}})}{2}+\\&-\ln{2}.
\end{aligned}
\end{equation} 
\normalsize 
Using now eq.(\ref{eq:uguaglianzacorretta}) and (\ref{eq:deltaScondFormal}) we obtain
\small
\begin{equation}\label{eq:deltaSconditional}
\begin{aligned}
\frac{\Delta S_{cond}}{K_b}=&-P_a^f\ln{P_a^f}-P_b^f\ln{P_b^f}+\\
&+\frac{1+(2P_a^f-1)\erf(\frac{h}{\sqrt{\pi}})}{2}\ln\frac{1+(2P_a^f-1)\erf(\frac{h}{\sqrt{\pi}})}{2}+\\ &+\frac{1+(2P_b^f-1)\erf(\frac{h}{\sqrt{\pi}})}{2}\ln\frac{1+(2P_b^f-1)\erf(\frac{h}{\sqrt{\pi}})}{2}+\\&-\frac{P_a^f}{\sqrt{2\pi}}\int_{-\infty}^\infty e^{-\frac{z^2}{2}}\ln\bigg(1+\frac{P_b^f}{P_a^f} e^{2zh-2h^2}\bigg)dz+\\
&-\frac{P_b^f}{\sqrt{2\pi}}\int_{-\infty}^\infty e^{-\frac{z^2}{2}}\ln\bigg(1+\frac{P_a^f}{P_b^f} e^{2zh-2h^2}\bigg)dz+\\
&+\frac{1}{\sqrt{2\pi}}\int_{-\infty}^\infty e^{-\frac{z^2}{2}}\ln\bigg(1+e^{2zh-2h^2}\bigg)dz.
\end{aligned}
\end{equation}
\normalsize
where the first three lines of the RHS are $\Delta S_{ex}$. 

\begin{figure*}
\includegraphics[width=0.9\linewidth]{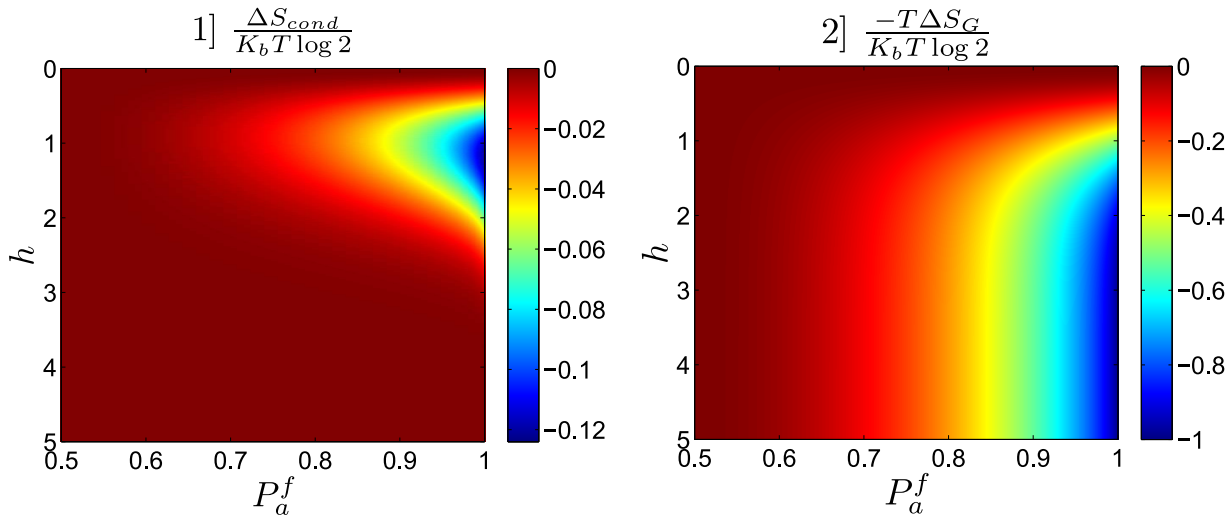}
\caption{Entropy variations for faulty reset protocols as functions of $P_a^f$ and $h$. The latter spans in $[0,5]$ because $h=5$ suffices to reach the asymptotic limit of small noise values (see eq.(\ref{eq:deltaSGibbs}) and (\ref{eq:deltaSconditional}) when $h\gg1$). In order: 1]$\Delta S_{cond}/(K_b\ln{2})$, 2]$\Delta S_{G}/(K_b\ln{2})$.}\label{fig:figure3}
\end{figure*}

Eq.(\ref{eq:deltaSGibbs}) and (\ref{eq:deltaSconditional}) are plotted in Fig. \ref{fig:figure3} as functions of $P_a^f$ and $h$. Figure \ref{fig:figure3}.1 shows our main result, that $\Delta S_{cond} \neq 0$ even for a system with a symmetric bistable potential $U(x)$. For this particular example, $\Delta S_{cond} \leq 0$ with $-0.12K_b\ln{2}$ as largest deviation from zero. Non-zero values are obtained for $h\lesssim1$ and $P_a^f\approx1$, a behaviour that can be explained in terms of eq.(\ref{eq:deltaScondFormal}), eq.(\ref{eq:sigmadef}) and eq.(\ref{eq:tedeltaU}). $P_a^f$ must be close to one because, for $P_a^f\approx0.5$, no reset takes place and entropy variations are zero. Moreover, for $h\lesssim 1$, thermal noise intensity $K_bT$ is comparable or greater than $\Delta U$ and $\sigma \gtrsim |x_a|$; this implies that each Gaussian peak has from $50\%$ to $85\%$ of its total area inside one logic subset and the remaining part in the other. In this case we see from eq.\ref{eq:formalProbabilities} that  $(P_0^f,P_1^f)\neq(P_a^f,P_b^f)$ by a significative amount. Thus, thermal noise limits the identification of a single logic state with a well defined Gaussian peak, providing $\Delta S_{ex} \neq 0$. The discussion for $\Delta S_{ov}\neq0$ is similar but less intuitive. When the noise is large ($h\approx1$), the two Gaussian peaks overlap within $1\sigma$. Peaks overlap is thus a relevant property of $P(x)$ and the integrals $I(\cdot)$ in eq.(\ref{eq:formalGibbs3}c) must be nonzero. Additionally, their sum in $S_{ov}$ strongly depends on $(P_a,P_b)$ specific values, so also $\Delta S_{ov}$ must be nonzero.

\begin{figure}
\includegraphics[width=\linewidth]{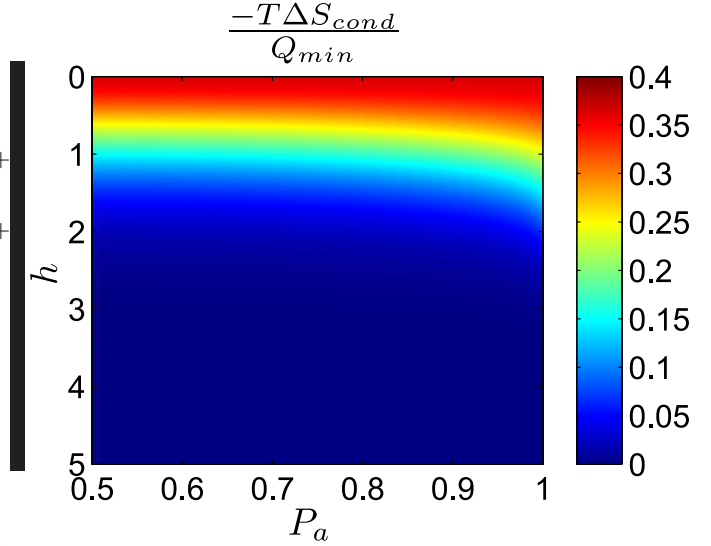}
\caption{The relative contribution of $-T\Delta S_{cond}$ to $Q_{min}$ as a function of $P_a^f$ and $h$. From this plot it's clear that $\Delta S_{cond}$ is non-negligible if $h\lesssim1$.}\label{fig:figure4}
\end{figure}

Since $\Delta S_{cond}$ can be nonzero for a reset protocol operated on a bistable symmetric system, we investigate its contributions to minimum heat production . This is given by 
\begin{equation}
Q_{min}=-T\Delta S_{G}=-T\Delta S_{S}-T\Delta S_{cond}
\end{equation}
where $-T\Delta S_{S}$ is the heat production for a reset with error probability $P_1^f$ \cite{gammaitoni,ciliberto} and $-T\Delta S_{cond}$ is the contribution due to conditional entropy. In our example we proved that $0\leq -T\Delta S_{G}\leq K_bT\ln{2}$ (Fig.\ref{fig:figure3}.2) and that $-T\Delta S_{S}\leq-T\Delta S_{G}$ (Fig.\ref{fig:figure3}.1). Putting everything together, we obtain 
\begin{equation}\label{eq:ineq_corr}
0\leq-T\Delta S_{S}\leq Q_{min} \leq K_bT\ln{2}.
\end{equation}
Two results are drawn from these inequalities. The first one is the well known $0\leq-T\Delta S_{S}\leq K_bT\ln{2}$ and states that error probabilities reduce the minimum heat produced to reset one bit \cite{ciliberto, gammaitoni}. The second is $-T\Delta S_{S}\leq Q_{min}$ and tells us that, in our example, minimum heat production is underestimated if $\Delta S_{cond}$ is improperly neglected. For a quantitative evaluation of the underestimation, see Fig. \ref{fig:figure4}

To conclude, we discuss the low noise limit $h\gg1$ (eq.(\ref{eq:tedeltaU})). From eq.(\ref{eq:finalProbabilities}), (\ref{eq:deltaSGibbs}), (\ref{eq:deltaSShannon}) and (\ref{eq:deltaSconditional}) we have:
\small
\begin{subequations}\label{eq:limit_results}
\begin{align}
&P_0^f\approx P_a^f, \quad P_1^f \approx P_b^f\\
&\Delta S_G\approx \Delta S_{cg}=-K_b(P_a^f\ln{P_a^f}+P_b^f\ln{P_b^f}+\ln{2})\\
&\Delta S_S\approx \Delta S_{cg}=-K_b(P_a^f\ln{P_a^f}+P_b^f\ln{P_b^f}+\ln{2})\\
&\Delta S_{cond}\approx0.
\end{align}
\end{subequations}
\normalsize
Putting eq.(\ref{eq:limit_results}a), (\ref{eq:limit_results}c) and (\ref{eq:limit_results}d) in eq.(\ref{eq:Clausiuscorretto}), we obtain
\begin{equation}\label{eq:heat}
Q\geq K_bT(P_0^f\ln{P_0^f}+P_1^f\ln{P_1^f}+\ln{2})
\end{equation}
which is the classical Landauer limit for bit-reset with error probability $P_1^f$ \cite{gammaitoni,ciliberto, shizume, piechocinska,dillenlutz}. This occurs because both the differences between $(P_a,P_b)$ and $(P_0,P_1)$ and the overlap integrals are given by Gaussian tails. For this reason they can be neglected and $\Delta S_{ex}\approx0$, $\Delta S_{ov}\approx0$ and $\Delta S_{cond}\approx 0$. \\
In the litterature \cite{gammaitoni,ciliberto, shizume, piechocinska,dillenlutz}, eq.(\ref{eq:heat}) is obtained with eq.(\ref{eq:thermaleqproto}) as nonequilibrium PDF.  In this paper we derive this result using weaker (and physically more reasonable) assumptions: eq.(\ref{eq:formalPDF}) as nonequilibrium PDF and $\Delta U/(K_bT)\gg1$. If we remove $\Delta U/(K_bT)\gg1$ assumption we obtain again eq.(\ref{eq:ineq_corr}), where the contribution of the conditional entropy to $Q_{min}$ becomes significant. As proved in the previous example, such contributions can't arise from eq.(\ref{eq:thermaleqproto}) if $\Delta U/(K_bT)\gg1$ is removed.

\section{V. Conclusions}

To study the role of conditional entropy in bistable symmetric systems, we proposed eq.(\ref{eq:formalPDF}) as a new way to represent nonequilibrium PDFs instead of the commonly used eq.(\ref{eq:thermaleqproto}). This more general PDF, which includes eq.(\ref{eq:thermaleqproto}) as special case, allowed us to show that conditional entropy plays a role also for symmetric bistable systems whereas the formulation based on eq.(\ref{eq:thermaleqproto}) always gives a null contribution. Thanks to this new formulation the conditional entropy can be written as the sum of three contributions directly and intuitively connected to the PDF structure.\\
To illustrate the aforementioned points, we discussed the reset of one bit with eq.\ref{eq:formalPDF} as nonequilibrium PDF. Here two scenarios are available. If $\Delta U/(K_bT)\gg1$, then $\Delta S_{cond}=0$ and known results on Landauer principle are obtained with a weaker set of assumptions. If $\Delta U/(K_bT)\lesssim1$, we obtain a new result: $\Delta S_{cond}$ is nonzero for bistable and symmetric systems and contributes significantly to the minimum heat production. We underline that the latter scenario is not of academical interest only: the scaling down trend in ICT is likely to produce a device that operate in the $\Delta U/(K_bT)\lesssim1$ regime within few years.

\acknowledgments
We acknowledge support by the European Union (FPVII (2007-20013) under G.A. n 318287 LANDAUER, G.A. n 270005 ZEROPOWER and G.A. n 611004 ICT-Energy).

\bibliographystyle{unsrt}
\bibliography{riferimenti}

\end{document}